\documentclass[%
 reprint,
superscriptaddress,
 amsmath,amssymb,
 aps,
 prl
]{revtex4-2}

\usepackage{graphicx}% Include figure files
\usepackage{dcolumn}% Align table columns on decimal point
\usepackage{bm}% bold math
\usepackage{xcolor}
\usepackage{graphicx}
    \graphicspath{{Manuscript Figures/}}
\begin{document}

\preprint{APS/123-QED}

\title{Inertia–Dilatancy Interplay Governs Shear-Thickening Drop Impact}% Force line breaks with \\

\author{Anahita Mobaseri}
\affiliation{\mbox{Department of Chemical Engineering and Materials Science, University of Minnesota, Minneapolis, MN 55455, USA}}
\author{Leonardo Gordillo}
\affiliation{\mbox{Departamento de F\'{i}sica, Facultad de Ciencia, Universidad de Santiago de Chile (USACH), Santiago, Chile
}} 
\author{Charles Burton}
\affiliation{\mbox{Department of Physics and Astronomy, Northwestern University, Evanston, IL 60208, USA}}
\author{Soyoon Yoon}
\author{Dong Lee}
\author{Satish Kumar}
\affiliation{\mbox{Department of Chemical Engineering and Materials Science, University of Minnesota, Minneapolis, MN 55455, USA}}
\author{Michelle M. Driscoll}
\affiliation{\mbox{Department of Physics and Astronomy, Northwestern University, Evanston, IL 60208, USA}}
\author{Xiang Cheng}
\email{xcheng@umn.edu}
\affiliation{\mbox{Department of Chemical Engineering and Materials Science, University of Minnesota, Minneapolis, MN 55455, USA}}
\affiliation{Saint Anthony Falls Laboratory, University of Minnesota, Minneapolis, MN 55414, USA}

\date{\today}% It is always \today, today,
             %  but any date may be explicitly specified
\begin{abstract}
Combining high-speed photography with direct force measurements, we investigate the impact dynamics of drops of cornstarch–water mixtures---a premier example of shear-thickening fluids---across a wide range of impact conditions. Our study identifies three distinct impact regimes. In addition to the liquid-like and solid-like behaviors generally expected for the impact-induced response of shear-thickening fluids, we uncover a counterintuitive regime in which high-concentration cornstarch–water mixtures display a liquid-like response at the onset of impact when shear rates are high and only transition to a solid-like behavior at later times as shear rates reduce. By integrating the classic drop-impact theory with the Reynolds–Darcy mechanism for dilatancy, we develop a unified model that quantitatively describes the impact dynamics of shear-thickening drops across all regimes. Our work reveals the unexpected response of shear-thickening fluids to ultra-fast deformation and advances fundamental understanding of drop impact for complex fluids.      
\end{abstract}

%\keywords{Suggested keywords}%Use showkeys class option if keyword
                              %display desired
\maketitle
From raindrops battering a windshield to coffee droplets splashing onto a table, everyday drop impact often belies the extreme mechanics underlying this ubiquitous fluid process \cite{Yarin2006_review,Josserand2016_review,Cheng2022_review}. Indeed, drop impact provides an elegant platform for probing fluid behavior under rapid deformation, inaccessible to standard rheometric techniques \cite{Macosko1994}. Motivated by this fundamental interest as well as its broad relevance in industrial, agricultural, and biomedical applications, the impact of drops of complex fluids with diverse rheological properties has been extensively studied \cite{Shah2024_review,mobaseri2025maximum}. Among these, shear-thickening fluids---most famously exemplified by cornstarch-water mixtures \cite{Brown2009_Cornstarch,Fall2012_cornstarch}---stand out due to their dramatic viscosity increase and striking shear jamming behavior at high shear rates \cite{Wagner2009Shearthickening,Cheng2011shearthickening,brown2014shear,Morris2020_review}, giving rise to rich and complex impact dynamics \cite{bertola2015impact, boyer2016drop, shah2022coexistence}. Nevertheless, compared with the in-depth studies on drop impacts of many other complex fluids and on the complementary problem of the response of bulk shear-thickening fluids under the impact of solid projectiles \cite{waitukaitis2012impact, crawford2013shear, von2013velocity, peters2014quasi, han2016high, jerome2016unifying, lim2017force, maharjan2018constitutive, allen2018system, mukhopadhyay2018testing, de2019high, lopez2020penetration}, our understanding of the impact of shear-thickening drops remains elementary, despite growing interest on shear-thickening drops in applications such as 3D printing \cite{ahmadzadeh2022fabrication} and soft robotics \cite{rus2015design}. 

Here, we study the impact dynamics of drops of cornstarch-water mixtures, subjected to a characteristic shear rate up to 2,000 s$^{-1}$. Unlike previous studies that solely focused on the kinematics of impacting drops using high-speed photography \cite{bertola2015impact, boyer2016drop, shah2022coexistence}, we directly measure the transient impact force of the drops, which reveals previously inaccessible dynamic features. Our combined kinematic and dynamic measurements challenge the prevailing view that a shear-thickening drop behaves like an elastic or viscoelastic sphere at high shear rates \cite{lopez2020penetration,shah2022coexistence,de2019high,maharjan2018constitutive,brown2014shear} and uncover an unexpected impact regime, where, despite extremely high shear rates, the drop shows a liquid-like response at short times, before transitioning to a solid-like behavior as the shear rates decrease over time. Our study provides new insights into the unusual behavior of shear-thickening fluids under ultra-fast deformation and offers a comprehensive understanding of the impact dynamics of shear-thickening drops, significantly advancing our knowledge of the impact of complex fluid drops.

We prepare suspensions of cornstarch (Carolina Biological Supply) in water with volume fractions $\phi$ ranging from $0.30$ to $0.43$, spanning rheological regimes from shear thinning to continuous and discontinuous shear thickening (Fig.~S1) \cite{supplemental}. Synchronized with a high-speed camera (Photron, SA-X2) that visualizes drop spreading, a force sensor (PCB, 209C11)  simultaneously measures the transient normal impact force, $F(t)$. For each experiment, a drop of a diameter of $D_0 = 4.0 \pm 0.2$ mm impacts an aluminum surface at a velocity, $U_0$, ranging from 0.5 up to 7.2 m/s. The density of the mixtures is $\rho = \rho_c \phi + \rho_w (1 - \phi)$, with the densities of cornstarch and water being $\rho_c = 1.55$ g/cm$^3$ and $\rho_{w} = 1.0$ g/cm$^3$.

\begin{figure*}
    \centering
    \includegraphics[width=0.9\linewidth]{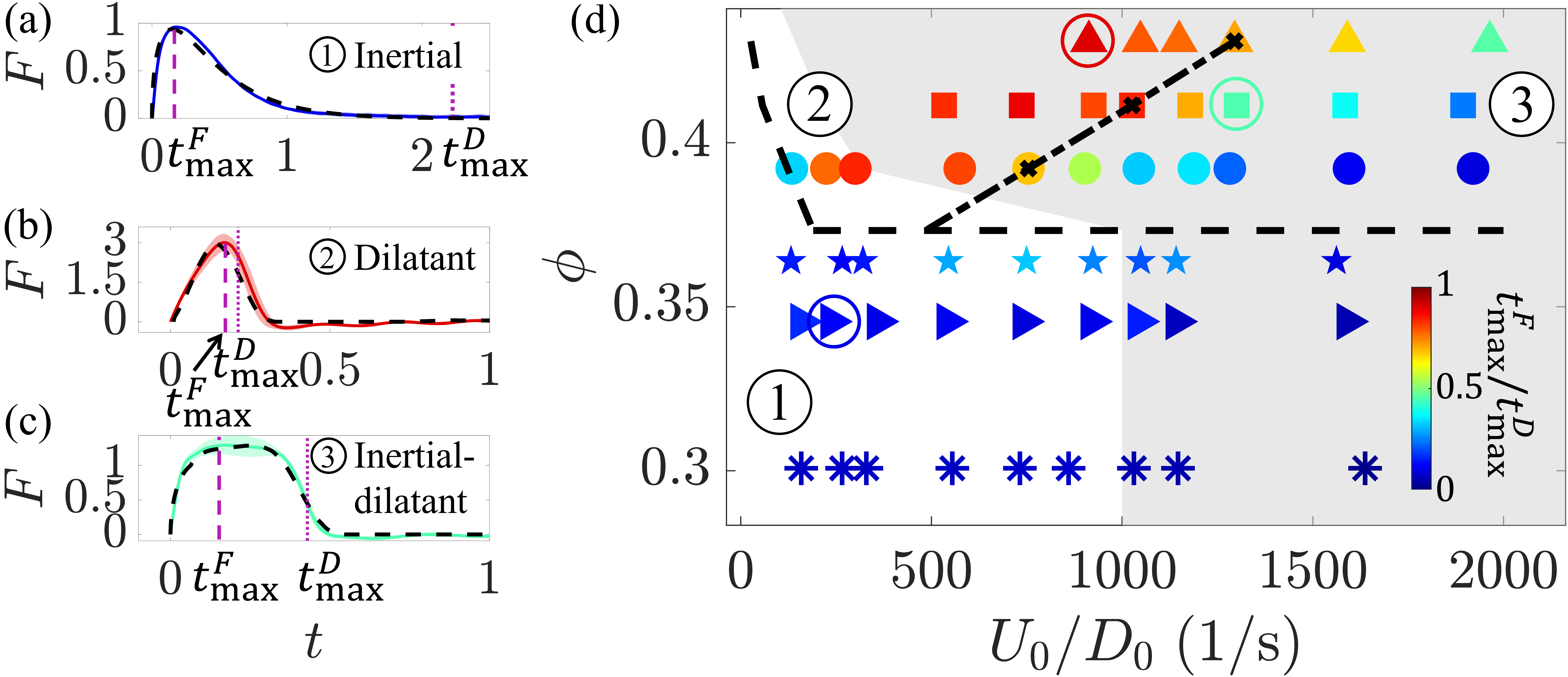}
    \caption{Phase diagram of the impact dynamics of shear-thickening drops.
    (a–c) Temporal evolution of the impact force, $F(t)$, for drops of cornstarch–water mixtures at: (a) volume fraction $\phi=0.35$, impact velocity $U_0 = 1.2$ m/s; 
    (b) $\phi=0.43$, $U_0 = 3.5$ m/s; and
    (c) $\phi=0.41$, $U_0 = 5.1$ m/s. Solid lines show averages over five experiments, with shaded regions indicating the standard deviation, while dashed lines correspond to model predictions. Vertical dashed and dotted lines indicate the time of maximum force $t_\text{max}^F$ and the time of maximum spreading $t_{\text{max}}^D$, respectively.  $F$ is normalized by $\rho U_0^2 D_0^2$, and time $t$ by $D_0/U_0$.
    (d) Ratio of the two times, $t_{\text{max}}^F / t_{\text{max}}^D$, plotted as a function of $\phi$ and characteristic shear rate $U_0/D_0$. Three regimes are indicated: \textcircled{1} Inertial, \textcircled{2} Dilatant, and \textcircled{3} Inertial–dilatant. Circled points correspond to the examples shown in (a–c). The dashed line shows the onset of discontinuous shear thickening, while the dot-dash line indicates the model prediction for the boundary between Regime (2) and (3), corresponding to $\text{ID} = 1$ (Eq.~\ref{eqn:eqn6}). Symbol shapes denote samples with different $\phi$ (the $y$-axis), and symbol colors indicate the value of $t_{\text{max}}^F / t_{\text{max}}^D$.
    The gray region marks the shear rates inaccessible to conventional rheometry. A three-dimensional representation of the dataset is provided in Fig.~S2 \cite{supplemental}.}
    \label{fig:fig1}
\end{figure*}

By examining the correlation between the kinematic and dynamic features of drops, we identify three distinct impact regimes. Specifically, we measure the ratio of the time of maximum force, $t_\text{max}^F$, to the time of maximum drop spreading, $t_\text{max}^D$, as a function of volume fraction $\phi$ and characteristic shear rate $U_0/D_0$ (Fig.~\ref{fig:fig1}(d)). \textit{Regime (1)} (Movie S1): At low volume fractions $\phi<0.37$ where mixtures exhibit continuous shear thickening, we observe $t_\text{max}^F/t_\text{max}^D \ll 1$ and a distinctly asymmetric impact force profile $F(t)$ about $t_\text{max}^F$ (Fig.~\ref{fig:fig1}(a)), both hallmarks of the inertia-driven impact of liquid drops \cite{gordillo2018dynamics, mitchell2019transient}. \textit{Regime (2)} (Movie S2): For $\phi>0.37$, mixtures exhibit discontinuous shear thickening at high shear rates. At relatively low impact velocity $U_0$---or equivalently, low shear rate $U_0/D_0$---the ratio $t_\text{max}^F/t_\text{max}^D$ approaches one (Fig.~\ref{fig:fig1}(d)), and the impact force $F(t)$ becomes more symmetric about $t_\text{max}^F$ (Fig.~\ref{fig:fig1}(b)), features typically associated with the impact of solid spheres \cite{falcon1998behavior}. \textit{Regime (3)} (Movie S3): For mixtures of high $\phi$, as the impact velocity and therefore the shear rate increase further, one might expect an even more pronounced solid-like response due to strong shear thickening at higher shear rates. Surprisingly, $t_\text{max}^F/t_\text{max}^D$ decreases in this third regime, reaching values comparable to those observed in liquid-drop impact (Fig.~\ref{fig:fig1}(d)). The force profile $F(t)$ exhibits new intriguing features (Fig.~\ref{fig:fig1}(c)): it rises sharply at early times, similar to liquid-drop impact, maintains a high plateau after reaching the peak force, and finally plummets to zero, reminiscent of solid-sphere impact.

To understand these multifaceted observations, we analyze the detailed impact dynamics and elucidate the physical mechanisms governing each of the three regimes.

\textit{Regime (1)} (small $\phi$): For low-$\phi$ mixtures showing continuous shear thickening (Fig.~S1), inertia dominates the impact process. The impact force is well described by the theory of liquid drop impact (Fig.~\ref{fig:fig1}(a)) \cite{gordillo2018dynamics,mitchell2019transient,Cheng2022_review}:
\begin{equation}
F(t) = cA_0\sqrt{t} \exp(-t/\tau)
\label{eqn:eqn1}
\end{equation}
where $c$ is a dimensionless constant, $A_0 = \rho U_0^{5/2}D_0^{3/2}$ is a dimensional factor, and $\tau$ represents the timescale at which the force decays at long times.

\begin{figure}  
    \centering
    \includegraphics[width=1\linewidth]{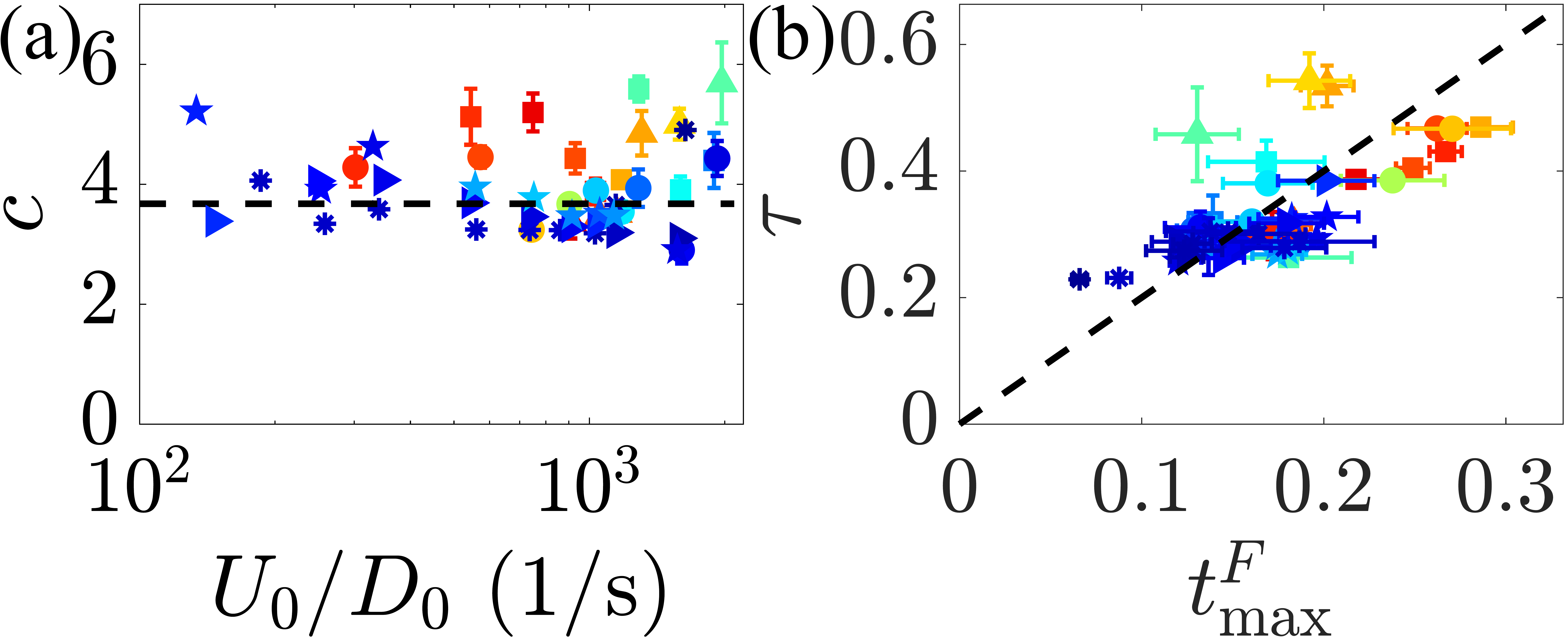}
    \caption{Liquid-drop impact behavior of shear-thickening fluids. (a) Prefactor $c$ in Eq.~\ref{eqn:eqn1} as a function of characteristic shear rate $U_0/D_0$. The dashed line shows $c = 3\sqrt{6}/2$. (b) Decay time $\tau$ in Eq.~\ref{eqn:eqn1} as a function of the time of maximum force $t_{\text{max}}^F$ extracted from experiments. The dashed line indicates $\tau=2t_{\text{max}}^F$. Times are normalized by $D_0/U_0$. Symbol shapes denote $\phi$, while colors indicate $t_{\text{max}}^F / t_{\text{max}}^D$, as in Fig.~\ref{fig:fig1}(d).
    } \label{fig:fig2}
\end{figure}

We extract $c$ and $\tau$ by fitting $F(t)$ under various impact conditions. Figure~\ref{fig:fig2}(a) shows $c$ as a function of the characteristic shear rate $U_0/D_0$. The data indicate an approximately constant $c = 3.99 \pm 0.75$, consistent with the theoretical prediction $c = 3\sqrt{6}/2 \approx 3.67$ \cite{gordillo2018dynamics}. Equation~1 further predicts a linear relation between the decay time and the time of maximum force, $\tau = 2 t_{\text{max}}^F$, which is again in quantitative agreement with experiments (Fig.~\ref{fig:fig2}(b)). A linear fit to the data yields $\tau = a\, t_{\text{max}}^F$ with $a = 1.88 \pm 0.05$. Hence, the impact of low-$\phi$ mixtures exhibiting continuous shear thickening can be approximated as the impact of Newtonian fluids driven by inertia \cite{mobaseri2025maximum}.

\textit{Regime (2)} (large $\phi$, low $U_0$): Previous studies on shear-thickening drop impact have largely focused on high-$\phi$ mixtures with discontinuous shear-thickening at low impact velocities due to the experimental challenge of achieving high impact velocities. Upon impact, a solidification front forms at the point of contact and propagates upward rapidly, abruptly arresting the drop's spreading and resulting in a maximum deformation that is insensitive to impact velocity \cite{bertola2015impact,boyer2016drop,shah2022coexistence}. As shear-thickening fluids that undergo shear jamming are often modeled as elastic or viscoelastic solids \cite{lopez2020penetration,shah2022coexistence,de2019high,maharjan2018constitutive,brown2014shear}, the Young’s modulus of impacting shear-thickening drops has been estimated accordingly \cite{shah2022coexistence}. The temporal variation of the impact force obtained in our experiments uncovers the details of the early-time dynamics that are not captured by drop kinematics alone, providing a more stringent test of the solid-response models. Fitting the measured $F(t)$ with elastic or viscoelastic models yields parameters (e.g., Young’s modulus and zero-shear viscosity) that are either internally inconsistent or in disagreement with independent measurements \cite{supplemental}. This finding suggests that, contrary to the conclusions of prior studies, an impacting drop undergoing discontinuous shear thickening cannot be modeled as a simple elastic or viscoelastic sphere.

The understanding of the early-time impact dynamics of Newtonian drops builds on the classical impact theory of Wagner \cite{philippi2016drop,gordillo2018dynamics}, originally developed to describe the water entry of solid bodies \cite{Wagner1932_impact}. The dual nature of solid-object impact on liquids and liquid-drop impact on solid substrates motivates us to apply the model of solid-sphere impact on a pool of cornstarch–water mixtures \cite{jerome2016unifying} to resolve the solid-like response of shear-thickening drops impacting flat solid substrates.

Above the critical volume fraction $\phi_c$, cornstarch-water mixtures show discontinuous shear thickening. The high shear rates induced by impact trigger Reynolds dilatancy of the cornstarch particles within the drop (Movie~S4) \cite{brown2014shear, bertola2015impact}, leading to the solid-like response observed in this regime. Specifically, dilatancy reduces the pore pressure, which drives Darcy flow through the interstitial pores and forces particles into closer contact. The process enhances interparticle friction and gives rise to the observed solid-like impact force. Quantitatively, the pore pressure $P_f$ is governed by the Reynolds–Darcy equation \cite{jerome2016unifying,supplemental}:  
\begin{equation}
    \nabla^2 P_f = \frac{\mu_w}{\kappa} \dot{\gamma} \tan\Psi = \frac{\mu_w}{\kappa} \dot{\gamma} \alpha (\phi - \phi_c),
    \label{eqn:eqn2}
\end{equation}
where $\dot{\gamma}$ is the shear rate, $\mu_w = 1$ mPa$\cdot$s is the water viscosity, and $\kappa = (1 - \phi)^3 a^2 / (180 \phi^2)$ is the permeability of the mixtures with $a \approx 10$ $\mu$m the average diameter of cornstarch particles. The extent of dilatancy is characterized by the Reynolds dilatancy angle $\Psi = \text{atan}(\alpha \Delta\phi)$, which scales with the deviation from the critical volume fraction, $\Delta\phi = \phi - \phi_c$, with a prefactor $\alpha \approx 1$ \cite{shi2021theoretical}.  Note that $\phi_c$ is constant at about $0.37$ at sufficiently high shear rates relevant to our study (Fig.~\ref{fig:fig1}(d)).

Using Eq.~\ref{eqn:eqn2}, the pore pressure can be estimated as 
\begin{equation}
    P_f \approx \frac{\mu_w}{\kappa} \dot{\gamma} \Delta \phi  h^2 \approx \frac{\mu_w}{\kappa} \Delta\phi r_c \left| \frac{dh}{dt} \right|,
     \label{eqn:eqn3}
\end{equation} 
% where $\dot{\gamma} = (d h/dt)/h$ is the shear rate and $\Delta h = D_0 - h(t)$ gives the length scale of deformation with $h(t)$ denoting the drop height at time $t$.
where the drop height at time $t$, $h(t)$, gives the deformation length scale. The characteristic shear rate $\dot{\gamma}$ is estimated as $u_r/h$, where the radial spreading velocity $u_r = \left|\frac{dr_c}{dt}\right| \approx \frac{r_c}{h} \left|\frac{dh}{dt}\right|$ and $r_c$ denotes the radius of the contact area \cite{supplemental,eggers2010drop}. The resulting contact stress $AP_f$ \cite{jerome2016unifying,Schofield1968} over the contact area $\pi r_c^2$ gives the impact force
\begin{equation}
    F(t) \approx A\pi r_c^3\frac{\mu_w}{\kappa} \Delta\phi \left|\frac{dh}{dt} \right|.
    \label{eqn:eqn4}
\end{equation}
Here, $A$ is the effective friction coefficient, which can be derived from the bearing capacity of a frictional granular bed and is related to the friction angle $\beta_0$ at $\phi_c$ by \cite{Schofield1968,jerome2016unifying} 
\begin{equation} \label{eqn:friction_coefficient}
    A = \frac{1-\sin\beta}{1+\sin\beta}\exp(\pi\tan\beta)
\end{equation} 
with $\tan\beta = \tan\beta_0+\tan\Psi$. Note that Laplace pressure from surface deformation is negligible and is therefore excluded from the calculation of $F(t)$ \cite{supplemental}.

Extracting $r_c(t)$ and $h(t)$ from the shape of the spreading drops, the Reynolds-Darcy model provides a quantitative description of the impact force in the solid-like regime (Fig.~\ref{fig:fig1}(b)). The fitting yields $\beta_0 = 44\pm 2^ \circ$, agreeing well with the repose angle of cornstarch powder \cite{yang2005dry}.

\textit{Regime (3)} (large $\phi$, high $U_0$): High-$\phi$ mixtures in this regime also exhibit discontinuous shear thickening, similar to those in Regime~(2), yet display a qualitatively different impact response. In particular, the mixtures show an unexpected re-entrant liquid-like behavior at high impact velocities, with $t_{\max}^F/t_{\max}^D$ dropping well below unity (Fig.~\ref{fig:fig1}(d)). The corresponding force profile has a hybrid character, appearing liquid-like at early times and solid-like at later times (Fig.~\ref{fig:fig1}(c)).

Below, we analyze the early and late stages of the impact separately to highlight the distinct dynamics that emerge over time.

\begin{figure}
    \centering
    \includegraphics[width=1\linewidth]{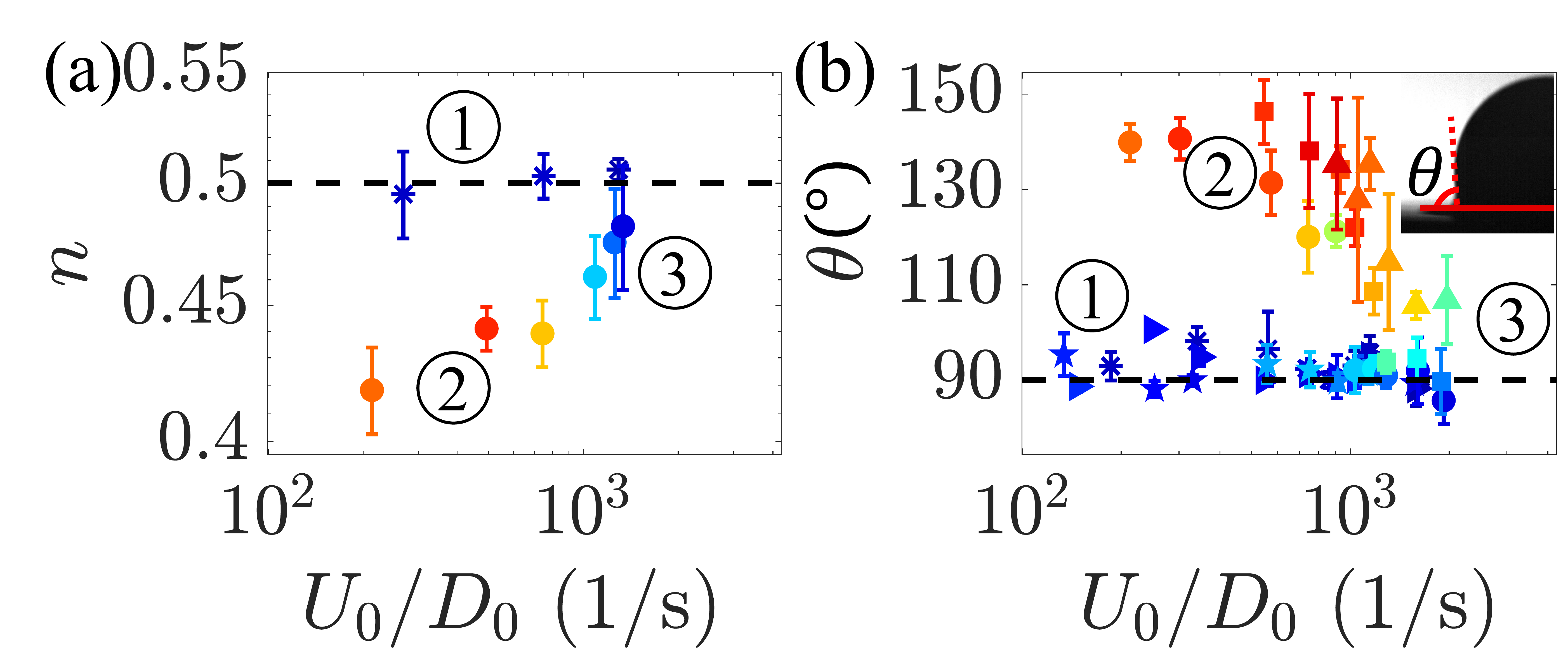}
    \caption{Early-time impact dynamics. (a) Spreading exponent $n$ as a function of the characteristic shear rate $U_0/D_0$. (b) Angle between the drop sidewalls and the impacted substrate, $\theta$, at the time of maximum force as a function of $U_0/D_0$. The dashed lines indicates the liquid-drop impact behaviors $n=0.5$ (a) and $\theta = 90^\circ$ (b). The inset shows $\theta$ for a $\phi=0.30$ sample. The three regimes are labeled. Symbol shapes denote $\phi$, while colors indicate $t_{\text{max}}^F / t_{\text{max}}^D$, as in Fig.~\ref{fig:fig1}(d).     
    }
    \label{fig:fig3}
\end{figure}

\underline{(\textit{i}) Early impact.} Before the peak force, we measure the spreading of the drop's contact line, $r(t)$, which follows a power-law behavior, $r(t)\sim t^n$ (Fig.~S3) \cite{supplemental}. In the liquid-like regime (Regime (1)), $n\approx0.5$ (Fig.~\ref{fig:fig3}(a)), consistent with the well-established result for inertia-driven liquid-drop impact \cite{philippi2016drop, gordillo2018dynamics, Sun2022_impactstress}. The exponent decreases to about 0.42 in the solid-like regime (Regime (2)), but returns to $n \approx 0.5$ in Regime (3), indicating a resurgence of inertial effects at early times.

In addition to the spreading dynamics, the timing of the peak force further underscores the dominant role of inertia at early times. In inertia-driven drop impact, the maximum force occurs when the drop's side walls are perpendicular to the surface \cite{mitchell2019transient, gordillo2018dynamics}, corresponding to the moment of the highest rate of change of vertical momentum. We measure the sidewall angle at the moment of peak force, $\theta(t=t_\text{max}^F)$, across all impact conditions (Fig.~\ref{fig:fig3}(b)). In the liquid-like regime (Regime (1)), $\theta(t_\text{max}^F) \approx 90^\circ$, as expected. In contrast, the angle consistently exceeds 
$90^\circ$ in the solid-like regime (Regime (2)), indicating that the peak force arises from a different, non-inertial mechanism. More importantly, $\theta(t_\text{max}^F)$ decreases and returns to $90^\circ$ in Regime (3), mirroring the trend observed in the spreading exponent $n$. 

Thus, both kinematic and dynamic measurements suggest that, although high-volume-fraction cornstarch–water mixtures exhibit strong discontinuous shear thickening, their early-time response remains inertia-dominated under high impact velocities, leading to a counterintuitive liquid-like behavior at the beginning of drop impact when the shear rate is extremely high.

\begin{figure}
    \centering
    \includegraphics[width=0.6\linewidth]{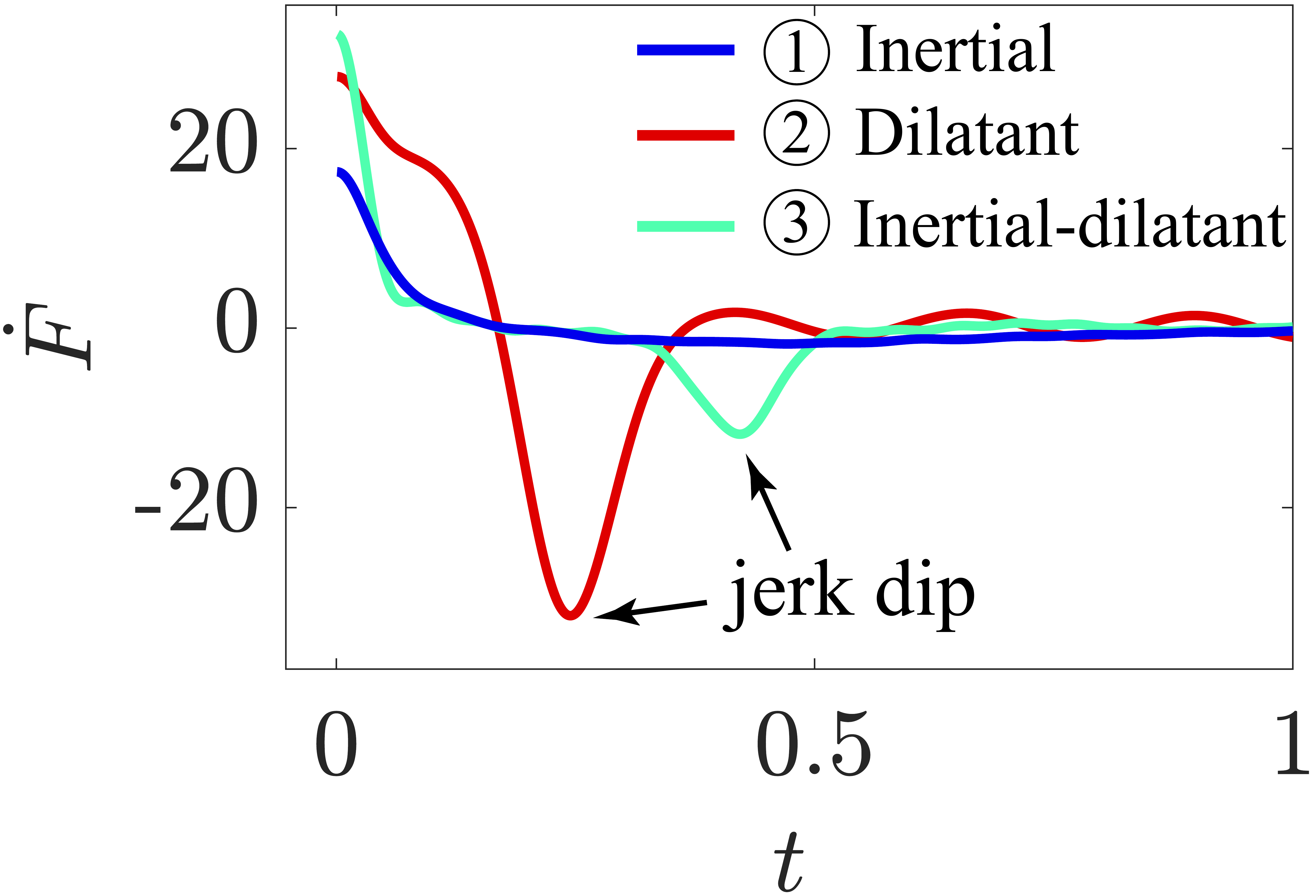}
    \caption{Late-time impact dynamics. Time derivative of the impact force, $\dot{F}$, for the three force curves in Fig.~\ref{fig:fig1}(a–c). $\dot{F}$ is non-dimensionalized by $\rho U_0^3D_0$ and $t$ by $D_0/U_0$.
    }
    \label{fig:fig4}
\end{figure}

\underline{(\textit{ii}) Late impact.} As the inertia of the drop decreases over time, the impact force induced by the Reynolds-Darcy mechanism gradually overtakes the inertial force. The solid-like response at late times in Regime (3) is evident when we examine the rate of change of force, $\dot{F}$, i.e., the jerk of the motion (Fig.~\ref{fig:fig4}).

A pronounced feature of the solid-like response is a sharp minimum in $\dot{F}$, indicating a rapid decline in the impact force at late times. This minimum, or `jerk dip', marks the complete transition of the impacting drop into a jammed solid (Fig.~S6) \cite{supplemental}, a feature absent in liquid-drop impact. The hybrid nature of the impact dynamics in Regime (3) is most clearly reflected in the behavior of $\dot{F}$ (Fig.~\ref{fig:fig4}): at early times, the curve matches that of a liquid-drop impact, showing a rapid decrease of the jerk, while at late times, it exhibits a pronounced negative minimum in $\dot{F}$, characteristic of solid-like response.

The physical picture outlined above allows us to demarcate the phase boundaries between the three impact regimes. First, because discontinuous shear thickening and the solid-like impact response originate from the same physical mechanism---namely Reynolds dilatancy \cite{brown2012role,brown2014shear}---the boundary between Regime~(1) and Regimes~(2) and~(3) is determined by the critical volume fraction $\phi_c \approx 0.37$ for the onset of discontinuous shear thickening, indicated by the dashed line in Fig.~\ref{fig:fig1}(d). Below $\phi_c$, cornstarch–water mixtures exhibit continuous shear thickening. In the absence of Reynolds dilatancy, drops composed of these low-$\phi$ mixtures display only inertia-driven, liquid-like impact responses. 

The transition between Regimes (2) and (3) is governed by the competition between inertia and dilatancy in the early stage of impact. While the inertial force scales as $F_i  \sim \rho U_0^2 D_0^2$, the dilatant force scales as 
$F_d \sim A\mu_wU_0{D_0}^3\Delta \phi/\kappa$ (Eq.~\ref{eqn:eqn4}), where we set $r_c \sim h \sim D_0$ and $dh/dt \sim U_0$. The ratio of these two forces, coined as the inertial–dilatant number (ID), is
\begin{equation}
    \text{ID} \equiv \frac{F_i}{F_d} = \frac{\rho \kappa U_0}{A \mu_w \Delta \phi D_0}.
    \label{eqn:eqn6}
\end{equation}
At large $\text{ID}$, inertia dominates the early-stage impact, giving rise to the counterintuitive reentrant liquid dynamics in Regime (3). Thus, $\text{ID} = 1$ marks the phase boundary between Regime (2) and (3) (the dash-dotted line in Fig.~\ref{fig:fig1}(d)). 

As the inertial timescale scales as $t_i\sim D_0/U_0$ \cite{gordillo2018dynamics,Cheng2022_review} and the timescale of dilatancy is given as $t_d \sim \rho \kappa/(A \mu_w \Delta \phi)$ \cite{supplemental,Strader_2022_timescale}, $\text{ID}$ can also be interpreted as the ratio of the dilatant to inertial times, $\text{ID}=t_d/t_i$. In Regime (2) with $\mathrm{ID} \ll 1$, dilatancy develops on a shorter timescale than inertia, resulting in a predominantly solid-like impact response at early times. In contrast, Regime~(3) corresponds to $\mathrm{ID} \gg 1$, where inertia dominates the early-time impact dynamics. Near the phase boundary with $\mathrm{ID} \sim 1$, the two timescales become comparable. The decay of one force is balanced by the growth of the other, giving rise to the extended force plateau observed in experiments. Indeed, we measure the duration of the maximum-force plateau, $\Delta t_p$, which exhibits a clear maximum near $\mathrm{ID} \sim 1$ (Fig.~S8), providing direct experimental support for the analysis.

\begin{figure}
    \centering
    \includegraphics[width=1\linewidth]{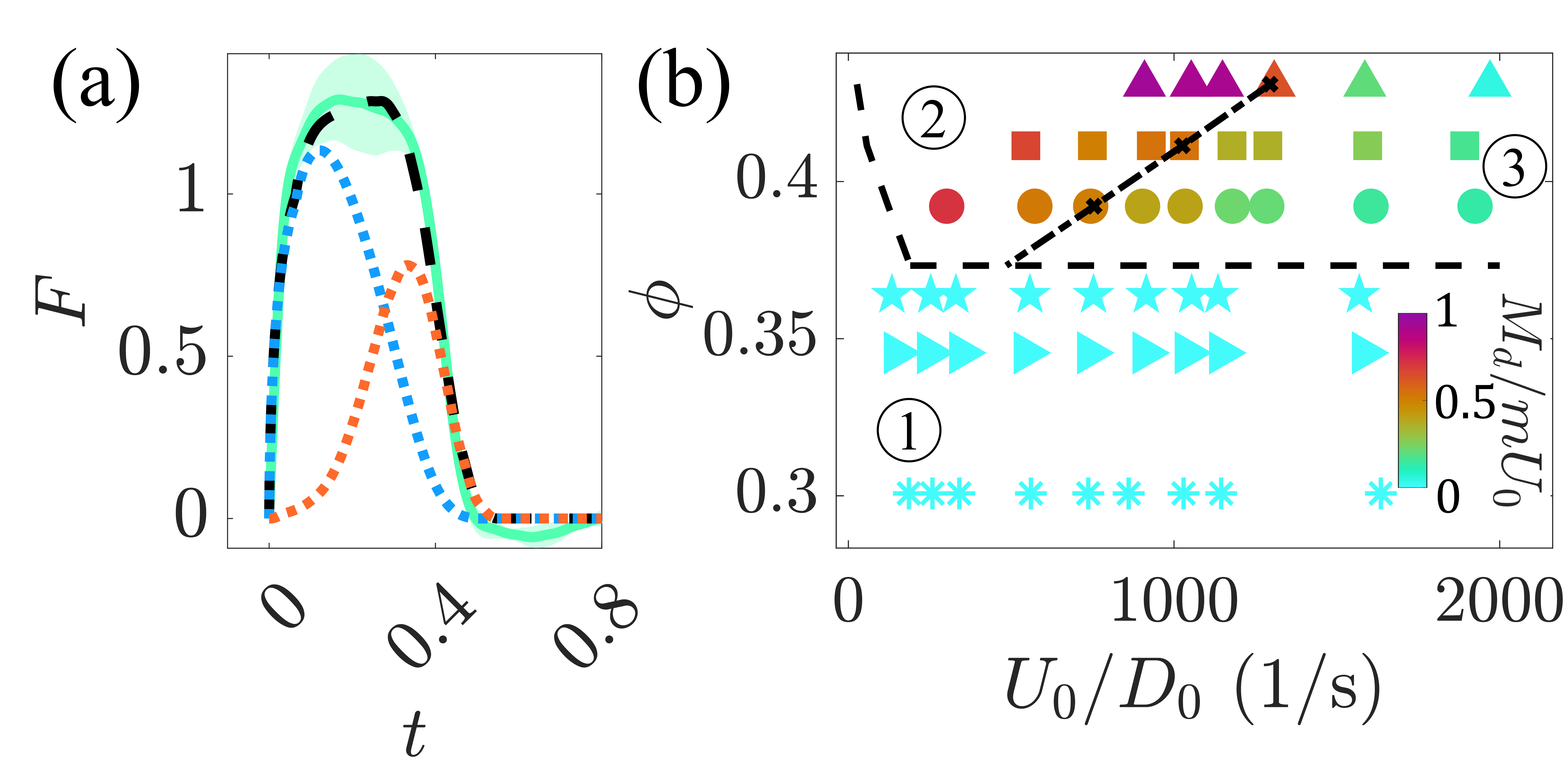}
    \caption{Unified force model. (a) Temporal evolution of impact force in Regime (3) for a volume fraction $\phi=0.41$ mixture at impact velocity $U_0 = 5.1$ m/s. The green solid line shows the average of five experiments, with the shaded region representing the standard deviation. The blue dotted line shows the weighted inertial impact force $(1-w_F)F_i$, the red dotted line shows the weighted dilatant impact force $w_FF_d$, and the black dashed line shows the summation of the two. Force is normalized by $\rho U_0^2D_0^2$ and time by $D_0/U_0$. (b) Ratio of the dilatant momentum $M_d$ to the total momentum $mU_0$ at different $\phi$ and $U_0/D_0$. Lines and symbol shapes are the same as those in Fig.~\ref{fig:fig1}(d). Symbol colors indicate $M_d/mU_0$.   
    }
    \label{fig:fig5}
\end{figure}

Guided by the scaling analysis above, we develop a unified model that quantitatively captures the impact force across all regimes. The relative contributions of inertial and dilatant stresses evolve over the course of a single impact event in Regime~(3), with the inertial response dominating at early times and the dilatant response dominating at late times. This observation necessitates the introduction of a time-dependent weighting function, $w_F(t)$, for describing the temporal evolution of the impact force. Physically, upon impact, the bottom portion of the drop solidifies first, forming a solidification front that propagates upward toward the top of the drop \cite{shah2022coexistence}. We show that the contribution of the dilatant stress to the total stress is proportional to the solidified fraction of the drop, thereby providing a physical basis for $w_F(t)$ (Sec.~7A in \cite{supplemental}). This theoretical result enables us to determine $w_F(t)$ experimentally by tracking the evolution of the solidification front within the drop (Fig.~S4). $w_F(t)$ increases from 0 at $t=0$ and approaches 1 at late times, when solidification is complete (Fig.~S5). The total impact force follows as $F(t) = \left[1 - w_F(t)\right] F_{i}(t) + w_F(t) F_{d}(t)$, where $F_i(t)$ and $F_d(t)$ are the inertial and dilatant impact forces from Eqs.~\ref{eqn:eqn1} and \ref{eqn:eqn4}, respectively. The simple model provides an excellent description of the highly structured impact force profile across regimes (see Fig.~\ref{fig:fig5}(a) for an example in Regime (3) and Figs.~S10–S15 for all datasets \cite{supplemental}). 

Figure~\ref{fig:fig5}(a) reveals that the reduction in inertial force is offset by a corresponding increase in dilatant force, producing the extended plateau around the peak force, as predicted by the scaling analysis. The fitting parameters $c$ and $\tau$ follow the same trend observed in Regime (1) (Fig.~\ref{fig:fig2}), and $\beta_0=41 \pm 5^\circ$ is consistent with that extracted from Regime (2).

%In Regime~(3), the prolonged coexistence of these two mechanisms gives rise to an extended force plateau, whose duration is maximized when inertial and dilatant forces are comparable, i.e., when the inertia–dilatant number, $ID$, is of order unity.
% Finally, we develop a unified model based on the above findings, which quantitatively captures the impact force across impact regimes \cite{supplemental}. Specifically, we obtain a time-dependent weighting function $w_F(t)$ that quantifies the fraction of the solidified region of the impacting drop at time $t$ by tracking the evolution of the solidification front within the drop (Figs.~S3 and S4) \cite{supplemental}. 
%\textcolor{red}{This paragraph needs to be updated once we addressed the issue on the timescales.} 

Our model enables quantitative assessment of the relative contributions of the solid-like and liquid-like responses to the conversion of the total impact momentum. The momentum change associated with the dilatancy-driven solid-like response can be calculated as $M_d = \int_0^\infty w_F(t)F_d(t) dt$, i.e., the area underneath the red dashed line in Fig.~\ref{fig:fig5}(a). The relative solid contribution to the momentum conversion is ${M_d}/mU_0$ (Fig.~\ref{fig:fig5}(b)), where $m=\pi\rho D_0^3/6$ is the drop mass. The conversion of drop momentum is dominated by the inertial force with $M_d/mU_0 = 0$ in Regime (1) and by the dilatant force with $M_d/mU_0 \approx 1$ in Regime (2). The analysis also shows that $M_d/mU_0$ decreases with increasing $U_0$, reaching approximately $0.5$ at the phase boundary between Regimes (2) and (3). These findings further confirm the existence of three distinct impact regimes (Fig.~\ref{fig:fig5}(b)).

By leveraging direct force measurements, our study uncovers the unusual dynamics of impacting shear-thickening drops that are inaccessible to high-speed photography. Force measurements are especially informative during the early stage of drop impact and in regimes where the drop exhibits solid-like behavior, critical for understanding the impact of shear-thickening drops. Deformation under these conditions is often too subtle to resolve optically yet produces pronounced force signals. 

The combined kinematic and dynamic measurements enable us to probe the behavior of shear-thickening fluids at ultra-high shear rates, opening a new avenue for impact-based rheometry. Particularly, our measurements reveal the co-existence of an inertia-driven liquid response and a dilatancy-driven solid response for high-volume-fraction cornstarch-water mixtures at high impact velocity. While our model assumes a simple separation of the liquid and solid regimes (Fig.~S9), captured by a time-dependent weighting function $w_F(t)$ \cite{supplemental}, it is likely that the actual separation is spatially more complex and better described by a spatiotemporal weighting function $w_F(\vec{r},t)$. Simulations that incorporate the dynamics of shear-thickening fluids and resolve the internal flow structure of the drop would help test this hypothesis.

%More broadly, our finding epitomizes the complexity of the fluid dynamics of drop impact and greatly enriches our understanding of the impact dynamics of non-Newtonian fluids. Insights into the response of impacting shear-thickening drops may inform the design of advanced materials for applications such as 3D printing \cite{ahmadzadeh2022fabrication}, protective gear \cite{liu2021acoustic} and soft robotics \cite{rus2015design}.   

\begin{acknowledgments}
We thank Vatsal Sanjay and Brian Seper for discussions, and Kimberly Kosto for access to a high-clearance lab space. The research was supported by NSF CBET-2505641 and DMR-2002817. We acknowledge general support from PPG Industries.
\end{acknowledgments}

\nocite{fall2015macroscopic,chatte2018shear,driscoll2011ultrafast,landau2012theory,hunter1957energy,reed1985energy,mclaskey2010hertzian,tanzi2019foundations,herrmann2013physics,jackson2000dynamics,brown2012role,loimer2002shear}

\bibliography{bib}% Produces the bibliography via BibTeX.

\end{document}